\documentclass[10pt,a4paper]{article}
\usepackage[utf8]{inputenc}
\usepackage{amsmath}
\usepackage{amsfonts}
\usepackage{amssymb}
\usepackage{graphicx}
\usepackage{xcolor}
\usepackage[square,numbers]{natbib}
\DeclareMathOperator*{\argmax}{arg\,max}
\author{Andrea Cappozzo\footnote{ Department of Statistics and Quantitative Methods, University of Milano-Bicocca, Milan, Italy, \texttt{andrea.cappozzo@unimib.it, francesca.greselin@unimib.it}}    \and
Ludovic Duponchel \footnote{Univ. Lille, CNRS, UMR 8516 - LASIRE–Laboratoire avancé de spectroscopie pour les interactions, la réactivité et l'environnement, F-59000 Lille, France, \texttt{ludovic.duponchel@univ-lille.fr}}  \and 
        Francesca Greselin$^*$ \and
        Thomas Brendan Murphy \footnote{School of Mathematics \& Statistics and Insight Research Centre, University College Dublin, Dublin, Ireland, \texttt{brendan.murphy@ucd.ie}}
}	 	 
\begin{document}

\title{Robust variable selection in the framework of
classification with label noise and outliers: applications to spectroscopic data in agri-food}
\date{\vspace{-5ex}}
\maketitle
\begin{abstract}
Classification of high-dimensional spectroscopic data is a common task in analytical chemistry. Well-established procedures like support vector machines (SVMs) and partial least squares discriminant analysis (PLS-DA) are the most common methods for tackling this supervised learning problem. Nonetheless, interpretation of these models remains sometimes difficult, and solutions based on feature selection are often adopted as they lead to the automatic identification of the most informative wavelengths. 
Unfortunately, for some delicate applications like food authenticity, mislabeled and adulterated spectra occur both in the calibration and/or validation sets, with dramatic effects on the model development, its prediction accuracy and robustness. 
Motivated by these issues, the present paper proposes a robust model-based method that simultaneously performs variable selection, outliers and label noise detection.
We demonstrate the effectiveness of our proposal in dealing with three agri-food spectroscopic studies, where several forms of perturbations are considered. Our approach succeeds in diminishing problem complexity, identifying anomalous spectra and attaining competitive predictive accuracy considering a very low number of selected wavelengths.
\end{abstract}
\section{Introduction}
Near-infrared (NIR) and mid-infrared (MIR) spectroscopy have nowadays become a standard analytical practice in countless fields, being fast and non-invasive techiniques for promptly characterizing samples of interest \cite{Pasquini2018,Valand2020a}. By acquiring a large number of absorbance values in a spectral range, NIR and MIR analyses provide compositional information for the products under study, with the final aim of being employed by chemometricians in developing multivariate models. Generally, variables in the so-obtained feature space appear in the order of thousands, undermining the usage of standard low-dimensional techniques \cite{Xiaobo2010}. To this extent, variable selection methods play a pivotal role in determining a relevant subset of wavelengths onto which perform any subsequent analysis \cite{Brown1992,Brenchley1997}. 
Indeed, the detection of the most informative segments in a spectral region offer numerous advantages. Firstly, it reduces problem complexity, leading to faster and more interpretable models. Secondly, loss on predictive power is generally avoided by excluding the contribution of irrelevant and redundant noisy areas. Thirdly, cost impact for future data collection and processing will be reduced. Fourthly, parameter estimation is qualitatively improved as a by product of the spectral selection. 
Lastly, spectral interpretation is facilitated, whence chemometricians may uncover previously unknown properties and differences among the considered samples \cite{Indahl2004}. For all the aforementioned reasons, chemometrics literature has always been greatly benefited by variable selection methodologies, and recent examples include the successful determination of soil properties \cite{Vohland2014}, yeast and oil concentration levels in beer and corn \cite{ZHAO2018}, yeast fermentation process using Raman spectroscopy \cite{Jiang2019}, holocellulose and lignin content in multispecies hardwoods \cite{Liang2020} and identification of adulterated Sanqi powder \cite{Chen2020}. 

Conceptually, a variable selection method requires a) the definition of a relevance measure and b) the choice of an algorithm to perform the search. 
Standard procedures used in chemometrics, such as variable influence on projection \cite{wold1993pls,eriksson2006multi}, selectivity ratio \cite{Rajalahti2009}, competitive adaptive reweighted sampling \cite{Li2009}, uninformative variable elimination \cite{Centner1996}, monte carlo-uninformative variable elimination \cite{Cai2008}, successive projections algorithm \cite{Araujo2001}, genetic algorithms \cite{Leardi1992},  sparse-based modeling \cite{Chun2010, Chung2010}, interval partial least squares \cite{Norgaard2000} and recursive weighted partial least squares \cite{Rinnan2014} fall within this quite general paradigm. Such methodologies are powerful techniques present in any chemometrician's data analysis toolbox, being off-the-shelf implementations readily available for most programming languages. As an example, the state-of-the-art \texttt{R} package \texttt{mdatools} includes routines for preprocessing, exploring and analyzing chemometrics data \cite{Kucheryavskiy2020}. Despite the well-established effectiveness of the above-mentioned methods, all their data-dependent steps rely on the implicit assumption that no corrupted samples are present in the dataset. 
That is, the employed relevance measures are not robust against noisy observations, so much so that, when adulterations occur, the reliability of the entire output may be jeopardized. Thankfully, spectroscopic data are most often recorded in controlled experiments. Nevertheless, there exist some delicate applications, such as sample authenticity in agri-food, in which the raw material itself may be spoiled and/or adulterated \cite{Reid2006}. Moreover, the experimental work-flow in spectroscopy generally encompasses diverse phases such as sample collection and preparation, spectral recording and, only afterwards, statistical analysis, with no guarantee that an end-to-end data quality control was performed: the existence of a small percentage of corrupted measurements and wrong labeling is the rule rather than the exception for most data acquisition processes \cite{Frenay2014}. Therefore, robust variable selection methods resistant to outliers and potential label noise are desirable. Particularly, the latter type of noise is seldom studied in analytical chemistry when developing a classification model, implicitly neglecting the circumstances in which such a situation may appear.  Spectra with low interclass and high intraclass variability, inadequacy of low-cost automatic labeling systems and/or inexperienced personnel, label inconsistency when multiple experts are tasked to classify the same sample, information loss and data-entry errors are only some of the causes that are likely to lead to mislabeling.


Motivated by the preceding arguments, the present article illustrates the capabilities of a robust variable selection method, recently introduced in the literature \cite{Cappozzo2020}, in performing high-dimensional classification in presence of label noise and outliers within a chemometrics context. The aim of the paper is thus to showcase how our procedure deals with contamination within the feature selection algorithm, without the need to resort to any data-cleaning preprocessing step.

Three successful applications to agri-food spectroscopic datasets will be analyzed and discussed. Specifically, the first part of the paper will briefly introduce the methodology and the datasets employed in the study. In the second part, model results will be presented, in comparison with state-of-the-art chemometric strategies. 
The manuscript concludes with a discussion, highlighting how the advantages of the proposed method could positively impact classification of samples from spectroscopic data.


\section{Material and methods}
\subsection{Robust variable selection: the stepwise REDDA approach} \label{sec:methodology}

The methodology described in the present Section falls within the model-based family of classifiers, coupled with a novel variable selection procedure resistant to outliers and label noise. The main concepts underlying the method are hereafter reported. 

Classification, also known as discriminant analysis, identifies the task of constructing a decision rule to assign an unlabeled sample to one of $G$ known classes. For doing so, a complete set of $N$ learning observations (i.e., the training set)
\begin{equation}
(\mathbf{x}, \mathbf{l})=\left\{\left(\mathbf{x}_{1}, \mathbf{l}_{1}\right), \ldots,\left(\mathbf{x}_{N}, \mathbf{l}_{N}\right) ; \mathbf{x}_{n} \in \mathbb{R}^{P},\,\, \mathbf{l}_{n}=\{l_{n1},\ldots, l_{nG}\}' \in \{0,1\}^G; \: n=1,\ldots,N\right\}
\end{equation}
is at our disposal; where $\mathbf{x}_{n}$ denotes a $P$-dimensional continuous predictor and $\mathbf{l}_{n}$ is its associated class label, such that $l_{ng}=1$ if observation $n$ belongs to group $g$ and $0$ otherwise with, clearly, $\sum_{g=1}^{G} l_{ng}=1 \: \forall n \in\{1,\ldots, N\}$. Specifically, in a spectroscopic dataset, $P$ represents the total number of spectral variables in which the absorbance value is recorded. 
Model-based classifiers 
require some probabilistic assumptions in terms of the data-generating mechanism: we assume that the prior probability of class $g$ is 
 $\tau_g>0$, $\sum_{g=1}^G\tau_g=1$. The $g$-th class-conditional densities are independent $P$-dimensional Gaussian, with mean vector $\boldsymbol{\mu}_g \in \mathbb{R}^P$ and $P \times P$ covariance matrix $\boldsymbol{\Sigma}_g$: $\mathbf{x}_n|\mathbf{l}_{ng}=1 \sim N_P(\boldsymbol{\mu}_g, \boldsymbol{\Sigma}_g)$. The joint density of $(\mathbf{x}_n, \mathbf{l}_n)$ is therefore given by:
\vspace{-0.2cm}
\begin{equation} \label{joint_density_EDDA_varsel}
p(\mathbf{x}_n,\mathbf{l}_n; \boldsymbol{\theta}) = p(\mathbf{l}_n;\boldsymbol{\tau})p(\mathbf{x}_n|\mathbf{l}_n; \boldsymbol{\mu}_g, \boldsymbol{\Sigma}_g)=\prod_{g=1}^G \left[ \tau_g \phi(\mathbf{x}_n; \boldsymbol{\mu}_g, \boldsymbol{\Sigma}_g) \right]^{l_{ng}}
\end{equation}
where $\phi(\cdot; \boldsymbol{\mu}_g, \boldsymbol{\Sigma}_g)$ denotes the multivariate normal density and $\boldsymbol{\theta}$ represents the collection of parameters to be estimated, $\boldsymbol{\theta}= \{ \tau_1, \ldots,  \tau_G, \boldsymbol{\mu}_1, \ldots, \boldsymbol{\mu}_G, \boldsymbol{\Sigma}_1, \ldots, \boldsymbol{\Sigma}_G \}$. Once the model has been fitted to the training set, test units $\mathbf{y}_m$, $m=1,\ldots,M$, are assigned to the $g$-th class via the maximum a posteriori (MAP) rule:

\begin{equation}
\argmax_{g \in \{1,\ldots,G\}}\frac{\hat{\tau}_{g} \phi\left(\mathbf{y}_{m} ; \hat{\boldsymbol{\mu}}_{g}, \hat{\boldsymbol{\Sigma}}_{g}\right)}{\sum_{j=1}^{G} \hat{\tau}_{j} \phi\left(\mathbf{y}_{m} ; \hat{\boldsymbol{\mu}}_{j}, \hat{\boldsymbol{\Sigma}}_{j}\right)}.
\end{equation}
This formulation identifies a quite generic supervised classification device, and its effectiveness in defining decision rules for spectroscopic datasets has been reported in \cite{Wu1996}, \cite{Dean2006}, \cite{Toher2007}, \cite{Murphy2012} and \cite{Jacques2010}, among others. For a general account on probabilistic model-based discriminant analysis and clustering methods in chemometrics, the reader is referred to the excellent review in \cite{Bouveyron2013}.

Among the many specifications developed from the probabilistic structure in \eqref{joint_density_EDDA_varsel}, the one considered here is the so called Eigenvalue Decomposition Discriminant Analysis (EDDA) \cite{Bensmail1996}. EDDA defines a family of constrained models, where different assumptions about the covariance matrices are imposed by considering the following eigenvalue decomposition:
\vspace{-0.2cm}
\begin{equation} \label{sigma_dec_varsel}
\boldsymbol{\Sigma}_g=\lambda_g\boldsymbol{D}_g\boldsymbol{A}_g\boldsymbol{D}^{'}_g
\end{equation}
where $\boldsymbol{D}_g$ is an orthogonal matrix of eigenvectors, determining groups orientation, $\boldsymbol{A}_g$ is a diagonal matrix such that $|\boldsymbol{A}_g|=1$, accounting for groups shape, and $\lambda_g=|\boldsymbol{\Sigma}_g|^{1/p}$ is a scalar that controls the associated volume. 
By imposing some of the quantities in \eqref{sigma_dec_varsel} to be equal across groups, the problem of over-parametrized modeling is mitigated. Notice that such class of
models is particularly flexible, as it includes very popular classification methods like Linear Discriminant Analysis (LDA) and Quadratic Discriminant Analysis (QDA) as special cases \cite{Hastie1996}. In details, LDA is obtained by constraining the covariance matrices to be group-wise equal, $\boldsymbol{\Sigma}_g=\boldsymbol{\Sigma}$, $\forall g \in \{1,\ldots,G\}$; whereas a QDA classifier is recovered when every quantity in \eqref{sigma_dec_varsel} is free to vary across components. REDDA \cite{Cappozzo2019b}, a robust model-based classifier, was introduced to extend the EDDA framework to handle label noise and outliers. REDDA 
is based on the maximization of a \textit{trimmed mixture log-likelihood} \cite{Neykov2007}, where a trimming level $\gamma$ assures that the most unlikely  $\lfloor N\gamma \rfloor$ data points under the postulated model are discarded, ultimately robustifying parameter estimates. Nonetheless, despite the parsimonious structure induced by the eigen-decomposition in \eqref{sigma_dec_varsel}, for analytical spectroscopic applications the number of variables can be much greater than the number of observations, so much so that the REDDA model may still suffer from the curse of dimensionality \cite{bellman1957dynamic}, jeopardizing its performance in high-dimensional spaces. To overcome this limitation, a recent contribution in the literature proposes to include a variable selection step within the REDDA framework \cite{Cappozzo2020}: the core methodology employed in the present paper. Under the reasonable assumption that only a portion of the spectral region is relevant for class discrimination, the procedure robustly identifies a subset of wavenumbers onto which building a (robust) decision rule. The attained output is a method that performs high-dimensional classification with variable selection, safeguarding it from potential label noise and outliers, identifying such anomalous samples as a by-product of the learning process.

The devised stepwise algorithm works as follows: we start from the empty set and, at each iteration, the inclusion of an extra variable into the model is evaluated, based on its robustly assessed discriminating power. In a similar fashion, the removal of an existing variable from the model is also considered. The procedure iterates between variable addition and removal until two consecutive steps have been rejected, then it stops. In details, at each iteration we partition the learning observations $\mathbf{x}_n$, $n=1,\ldots,N$, into three parts $\mathbf{x}_n=(\mathbf{x}_n^{c},x_n^{p},\mathbf{x}_n^{o})$, where:
\begin{itemize}
\item $\mathbf{x}_n^{c}$ indicates the set of variables currently included in the model
\item $x_n^{p}$ the variable proposed for inclusion
\item $\mathbf{x}_n^{o}$ the remaining variables.
\end{itemize}
The intent here is to determine whether $x_n^{p}$ shall be included (excluded) into (from) the relevant subset. To do so, we recast the problem as a model selection task, comparing the following two competing models:
\begin{figure}[!htbp]
\centering
\vspace*{-4.3cm}
\includegraphics[scale=.9]{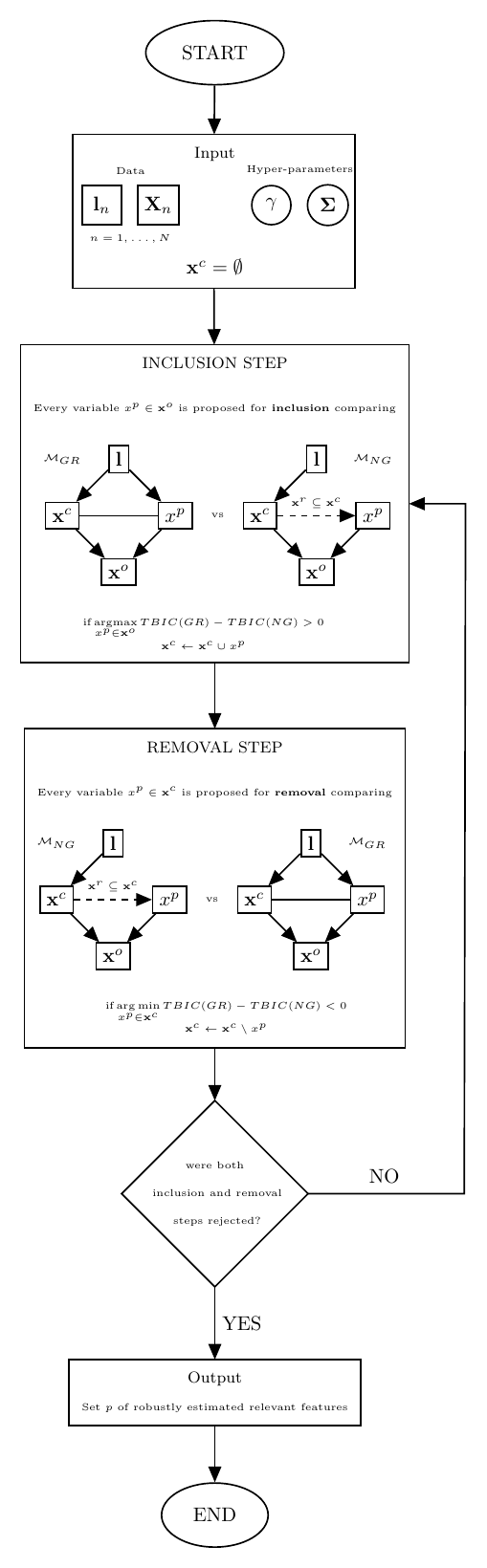}
\vspace*{-.7cm}
\caption{Flowchart of the stepwise REDDA algorithm. $\boldsymbol{\Sigma}$ denotes the covariance structure to be chosen within the available constrained models, according to Equation \eqref{sigma_dec_varsel} \cite{Bensmail1996}.}
\label{fig:var_sel}
\end{figure}
\begin{itemize}
\item \textit{Grouping ($\mathcal{M}_{GR}$):} $p(\mathbf{x}_n|\mathbf{l}_n)=p(\mathbf{x}_n^{c},x_n^{p},\mathbf{x}_n^{o}|\mathbf{l}_n)=
p(\mathbf{x}_n^{c},x_n^{p}|\mathbf{l}_n)p(\mathbf{x}_n^{o}|x_n^{p},\mathbf{x}_n^{c})$
\item \textit{No Grouping ($\mathcal{M}_{NG}$):} $p(\mathbf{x}_n|\mathbf{l}_n)=p(\mathbf{x}_n^{c},x_n^{p},\mathbf{x}_n^{o}|\mathbf{l}_n)=p(\mathbf{x}_n^{c}|\mathbf{l}_n)
p(x_n^{p}|\mathbf{x}_n^{r} \subseteq \mathbf{x}_n^{c})p(\mathbf{x}_n^{o}|x_n^{p},\mathbf{x}_n^{c})$
\end{itemize}
where $\mathbf{x}_n^{r}$ denotes a subset of the currently included variables $\mathbf{x}_n^{c}$. Model $\mathcal{M}_{GR}$ assumes that $x_n^{p}$ provides extra grouping information beyond that provided by $\mathbf{x}_n^{c}$; whereas $\mathcal{M}_{NG}$ specifies that $x_n^{p}$ is conditionally independent of the group membership given $\mathbf{x}_n^{r}$. We consider $\mathbf{x}_n^{r}$ in the conditional distribution because $x_n^{p}$ might be related to only a subset of the grouping variables $\mathbf{x}_n^{c}$ \cite{Maugis2011}. According to the general model-based structure described at the beginning of the Section, we assume  $p(\mathbf{x}_n^{c},x_n^{p}|\mathbf{l}_n)$ and $p(\mathbf{x}_n^{c}|\mathbf{l}_n)$ to be normal densities with constrained covariances depending on the class label $\mathbf{l}_n$, while $p(\mathbf{x}_n^{o}|x_n^{p},\mathbf{x}_n^{c})$ is required to be the same for both grouping and no grouping specifications. Exploiting standard results for multivariate normal theory, (see, for example, Theorem 3.2.4 in \cite{bibby1979multivariate}) $p(x_n^{p}|\mathbf{x}_n^{r} \subseteq \mathbf{x}_n^{c})$ defines a linear regression model. 
More specifically, the competing models are of the form:
\begin{align*}
\begin{split}
\mathcal{M}_{GR}: \left( \mathbf{x}_n^{c},x_n^{p}\right),\mathbf{l}_n \sim \prod_{g=1}^G \left[ \tau_g^{cp} \mathcal{N} (\mathbf{x}_n; \boldsymbol{\mu}_g^{cp}, \boldsymbol{\Sigma}_g^{cp}) \right]^{l_{ng}}\\
\mathcal{M}_{NG}: \mathbf{x}_n^{c},\mathbf{l}_n \sim \prod_{g=1}^G \left[ \tau_g^{c} \mathcal{N} (\mathbf{x}_n; \boldsymbol{\mu}_g^{c}, \boldsymbol{\Sigma}_g^{c}) \right]^{l_{ng}}, \quad  x_n^p|\mathbf{x}_n^{r} \sim \mathcal{N} \left( \alpha + \boldsymbol{\beta}^{'} \mathbf{x}_n^{r}, \sigma^2\right).
\end{split}
\end{align*}
A standard way to perform model comparison is via the Bayes Factor ($\mathcal{B}_{GR,NG}$) \cite{Kass1995}, evaluating the plausibility of the \textit{Grouping} model with respect to the \textit{No Grouping} one. In the stepwise REDDA approach, a robust proxy to $\mathcal{B}_{GR,NG}$ is used to select which specification to prefer. Following \cite{Raftery},  twice the logarithm of $\mathcal{B}_{GR,NG}$ is approximated with

\begin{equation} \label{tbic_diff}
2\log{\left(\mathcal{B}_{GR, NG}\right)}\approx TBIC(GR)-TBIC(NG)
\end{equation}
where the trimmed BIC (TBIC), firstly introduced in \cite{Neykov2007}, acts as a robust version of the Bayesian Information Criterion \cite{Schwarz1978} employed in the approximation in \eqref{tbic_diff}. Particularly, for the \textit{Grouping} and the \textit{No Grouping} specification outlined above, the TBICs respectively read:

\begin{align} \label{TBIC_gr}
\begin{split}
TBIC(GR) &= \underbrace{2\sum_{n=1}^N \zeta(\mathbf{x}_n^{c}, x_n^p)\sum_{g=1}^G l_{ng} \log{\left(\hat{\tau}_g^{cp} \phi(\mathbf{x}_n^{c}, x_n^p; \hat{\boldsymbol{\mu}}_g^{cp}, \hat{\boldsymbol{\Sigma}}_g^{cp})\right)}}_{2 \times \text{trimmed log maximized likelihood of }p(\mathbf{x}_n^{c},x_n^{p},\mathbf{l}_n)}
+ \\
&- v^{cp} log(N^*)
\end{split}
\end{align}
\begin{align} \label{TBIC_no}
\begin{split}
TBIC(NG) &= \underbrace{2\sum_{n=1}^N \iota(\mathbf{x}_n^{c}, x_n^p)\sum_{g=1}^G l_{ng} \log{\left(\hat{\tau}_g^{c} \phi(\mathbf{x}^c_n; \hat{\boldsymbol{\mu}}_g^{c}, \hat{\boldsymbol{\Sigma}}_g^{c})\right)}}_{2 \times \text{trimmed log maximized likelihood of }p(\mathbf{x}_n^{c}, \mathbf{l}_n)} - v^{c} log(N^*)+\\
&\underbrace{+2\sum_{n=1}^N \iota(\mathbf{x}_n^{c}, x_n^p) \log{\left[\phi\left(x_n^{p}; \hat{\alpha} + \hat{\boldsymbol{\beta}}^{'}\mathbf{x}_n^{r},\hat{\sigma}^2\right)\right]}}_{2 \times \text{trimmed log maximized likelihood of }p(x_n^{p}|\mathbf{x}_n^{r} \subseteq \mathbf{x}_n^{c})} - v^{p}log(N^*).
\end{split}
\end{align}
The quantities $v^{cp}$ and $v^{c}$ are penalty terms, namely the number of parameters for a REDDA model estimated on the set of variables $\mathbf{x}_n^c, x_n^p$ and $\mathbf{x}_n^c $, respectively; while $v^{p}$ accounts for the number of parameters in the linear regression of $x_n^{p}$ on $\mathbf{x}_n^{r}$. The terms $\zeta(\cdot)$ and $\iota(\cdot)$ are 0-1 indicator functions, identifying the subset of observations that have null weight in the trimmed likelihood under $\mathcal{M}_{GR}$ and $\mathcal{M}_{NG}$, with $N^*=\sum_{n=1}^N \zeta(\mathbf{x}_n)=\sum_{n=1}^N \iota(\mathbf{x}_n)$. That is, potential outlying and mislabeled observations do not influence the selection procedure, since only the most plausible $N^*=\lceil N(1-\gamma) \rceil$ samples are accounted for parameters estimation, with $\gamma$ denoting the impartial trimming level. The set of parameters $\left\{\alpha, \: \boldsymbol{\beta}, \: \sigma^2  \right\}$ are related to the linear regression component, and are robustly estimated via maximum likelihood on the untrimmed samples. 

In the addition stage, the $x_n^p$ variable with highest positive difference in $\eqref{tbic_diff}$ (if any) is the one selected for inclusion. In the removal stage, $x_n^p$ takes the role of the variable to be dropped, and the one displaying highest negative difference in $\eqref{tbic_diff}$ (if any) is excluded from the set of currently included variables $\mathbf{x}_n^{c}$. When neither addition nor removal move is performed, the procedure terminates. In this way, the number of relevant variables necessary to build the classification rule is automatically inferred, and it needs not be a priori specified. For those readers not interested in the mathematical details, a visual representation of the algorithm is reported in Figure \ref{fig:var_sel}, where a flowchart describes the main steps involved in the feature selection process.

Compared to other stepwise variable selection strategies developed for discriminant classification, such as stepwise LDA \cite{McCabe1975}, the obtained extension is twofold. Firstly, and most importantly, stepwise REDDA performs a robust search
protected from the harmful influence of label noise and outliers thanks to impartial trimming. 
Secondly, as previously mentioned, more flexibility in the covariance structure can be achieved via the decomposition in (4)  That is, stepwise LDA can be retrieved setting $\boldsymbol{\Sigma}_g=\boldsymbol{\Sigma}$ $\forall g$ and $\gamma=0$. The routines for the stepwise REDDA approach have been written in \texttt{R} language \cite{RCoreTeam2020}: the source code is openly available at \texttt{https://github.com/AndreaCappozzo/varselTBIC}.

\subsection{Spectroscopic datasets in agri-food}
The stepwise REDDA approach is applied to the analysis of three different multi-class datasets. The first one is a 4-class problem where the observations are mid-infrared spectra of modified starches. The second one is a 5-class problem encompassing visible and near infrared reflectance spectra of homogenized meat samples. The last dataset is a 2-class problem, concerning the discrimination between Ligurian and Non-Ligurian olive oil through mid-infrared measurements. For the considered datasets, employed instrumentation and sample collection procedures are thoroughly described in \cite{FernandezPierna2005}, \cite{McElhinney1999} and \cite{Hennessy2009}; thus, only a succinct explanation will be hereafter reported. All these challenging situations represent typical high-dimensional classification tasks often encountered in spectroscopy: a variable selection procedure resistant to label noise and outliers can potentially be beneficial in this regard. As a last worthy note, we mention that the subsequent analyses are directly performed on the raw spectra, without any pretreatment applied to the samples originally provided. Undeniably, adjacent wavelengths virtually contain the same information in terms of class separation; this is especially true for Near-infrared spectra \cite{Indahl2004}. Therefore, an option for reducing the computational complexity of the search would be to firstly group wavenumbers in pseudo-bands, and to subsequently look for the most discriminative spectral portions. This procedure may however introduce two potential shortcomings. On the one hand, the classification error could increase as soon as any adjacent wavelengths are aggregated, see, e.g., Table 9 in \cite{Murphy2012}. On the other hand, outlying spectra may be masked in the process. For these reasons, we decide not to consider wavelengths aggregation in the subsequent analyses.

\subsubsection*{Starches dataset}
\begin{figure}
\includegraphics[scale=.55]{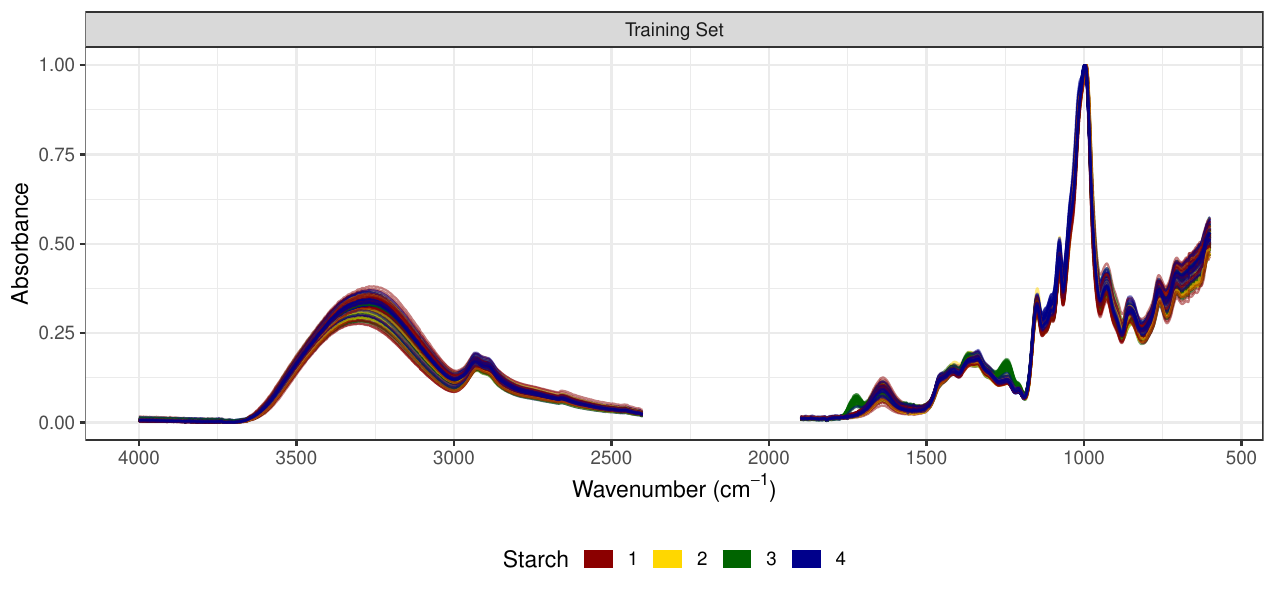}
\centering
\caption{Starches dataset: mid-infrared spectra of four starches classes}
\label{fig:starches_training}
 \end{figure}
\begin{figure}
\includegraphics[scale=.4]{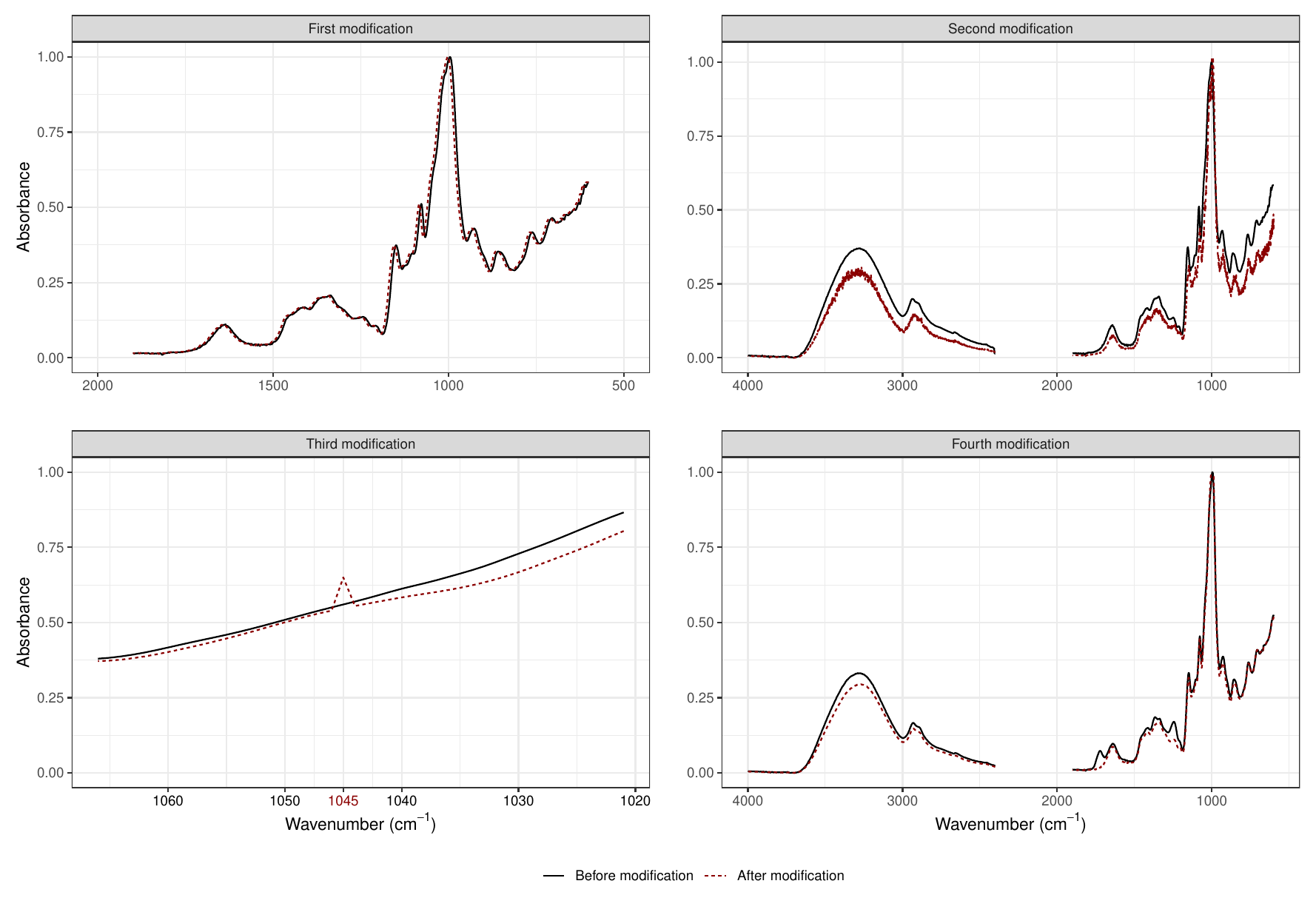}
\centering
\caption{Starches dataset: the 4 adulterated spectra manually placed in the test set by the `Chimiom\'{e}trie 2005' contest organizers, before and after modification.}
\label{fig:starches_modifications}
 \end{figure}

The first dataset comes from the chemometric challenge organized during the `Chimiom\'{e}trie 2005' conference \cite{FernandezPierna2007a}. The learning scenario encompasses $N=215$ training and $M=43$ test MIR spectra of starches of $G=4$ different classes. For each sample, a total of $P=2901$ absorbance measurements are recorded. A subset of training observations is displayed in Figure \ref{fig:starches_training}. The participants of the competition were tasked to discriminate as accurately as possible the four different classes, defining a classification rule from the training set. In addition, outlier detection needed to be performed, as four intentionally corrupted spectra were manually placed in the test set: a graphical representation is depicted in Figure \ref{fig:starches_modifications}. For a thorough description on how these modifications were obtained, the interested reader is referred to \cite{FernandezPierna2007a}. In addition, we slightly complicated the learning framework including less than $2\%$ of label noise in the training set: the last four samples of the third class were wrongly labeled as coming from the fourth one. 

\subsubsection*{Meat dataset}

The second dataset reports the NIR spectra of $231$ homogenized meat samples acquired in reflection mode from
$400$ to $2498$ nm at intervals of $2$ nm, accounting for a total of $P=1050$ spectral variables. Spectra belong to five different meat types, with $32$ beef, $55$ chicken, $34$ lamb, $55$ pork, and $55$ turkey. Data are collected using a Foss NIRSystems 6500 instrument which has two sensors: one for $400-1100$ nm and one for $1100-2498$ nm; this causes the discontinuity visible at wavenumber $1100$ in Figure \ref{fig:meat_training}.
We randomly partition the recorded spectra into test and calibration sets: the former is composed by $16$ beef, $27$ chicken, $17$ lamb, $27$ pork and $27$ turkey, 
while the latter contains the same proportion of these five meat types with four additional spectra manually adulterated as follows:
\begin{figure}
\includegraphics[scale=.55]{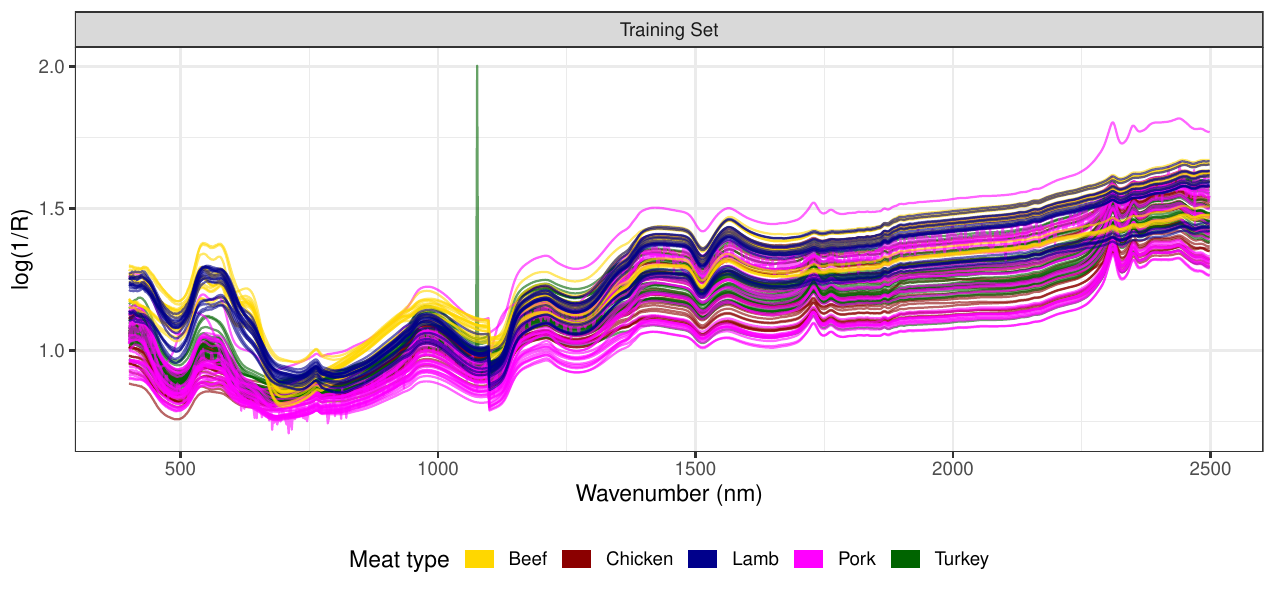}
\centering
\caption{Meat dataset: visible and near infrared spectra of five homogenized meat types. Discontinuity at $1100$ nm is due to a change in detector at that wavelength. Training set with $4$ outlying spectra.}
\label{fig:meat_training}
 \end{figure}
\begin{itemize}
\item a shifted version of a pork meat spectrum, achieved by removing the first $15$ data points and appending the last $15$ group-mean absorbance values at the end of it;
\item a noisy version of a pork meat spectrum, generated by adding Gaussian white noise to the original one;
\item a modified version of a turkey meat spectrum, obtained by abnormally increasing the absorbance value in a single specific wavelength to simulate a spike;
\item a pork meat spectrum with an added slope, produced by multiplying the original spectrum by a positive constant.
\end{itemize}
These modifications mimic the ones considered in the ``Chimiom\'{e}trie 2005''  chemometric contest for the starches dataset, described in the previous Section, and agree with those reported in \cite{Denti2020} within a novelty detection framework.

\subsubsection*{Olive Oil dataset}
\begin{figure}
\includegraphics[scale=.55]{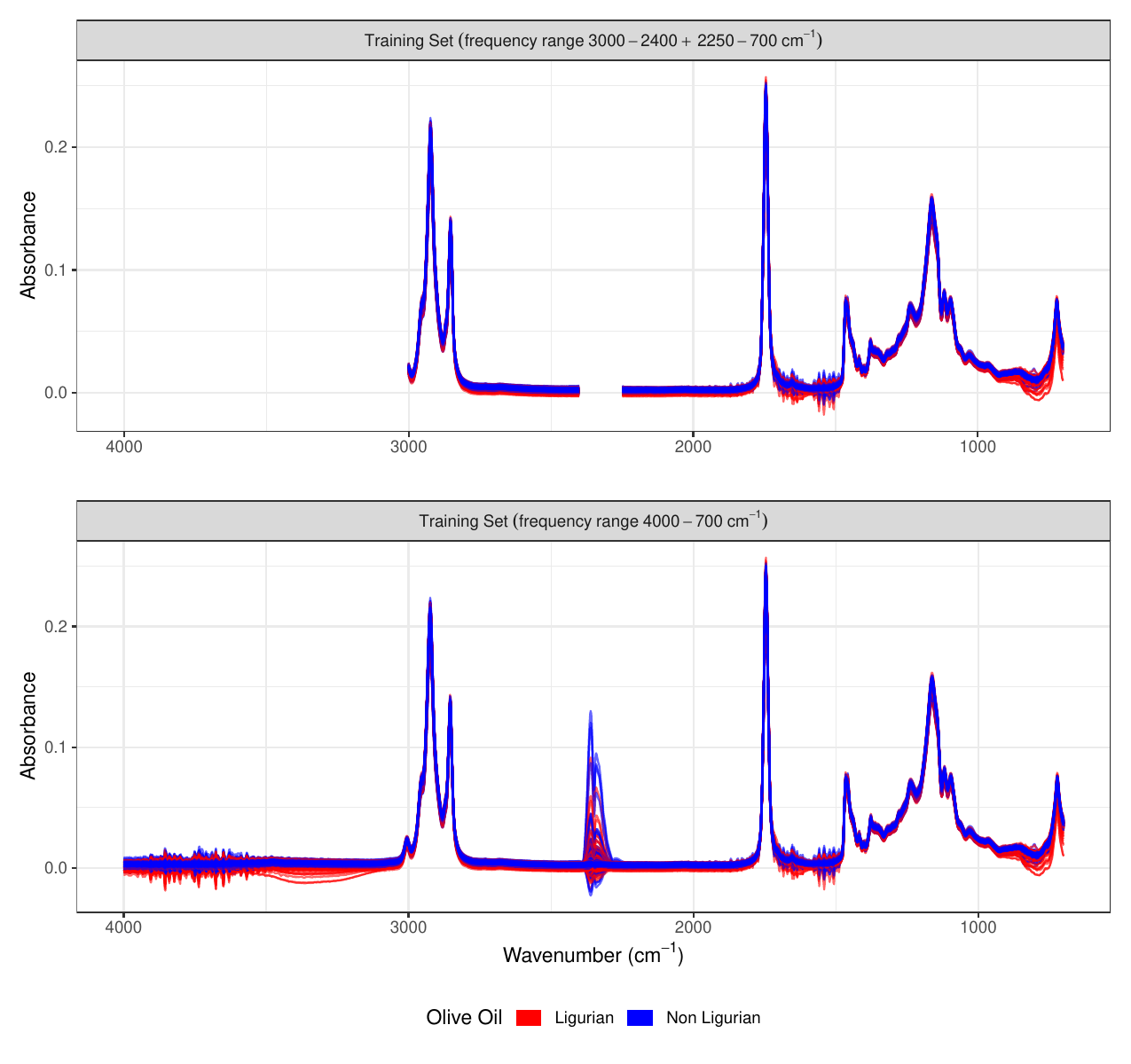}
\centering
\caption{Olive Oil dataset: mid-infrared spectra of Ligurian and Non Ligurian olive oil, spectral zones $3000-2400$ cm$^{-1}$ + $2250-700$ cm$^{-1}$ (top panel) and $4000-700$ cm$^{-1}$ (bottom panel). }
\label{fig:FTIR_training}
 \end{figure}

The last dataset examines MIR olive oil spectra based on Fourier-transform
(FTIR) measurements considering attenuated total reflectance, with a learning scenario encompassing training and test sets of sizes respectively equal to $N=280$ and $M=630$. The aim here is to identify whether or not samples originate from the Italian coastal region of Liguria.  In doing so, two nested subsets of wavelengths are considered in the analysis: the first one comprises the spectral zones from $3000$ to $2400$ cm$^{-1}$ and from $2250$ to $700$ cm$^{-1}$ ($P=1117$ recorded features), while the second one covers the entire $4000-700$ cm$^{-1}$ spectral range ($P=1712$). The two resulting training sets are respectively displayed in the top and bottom panels of Figure \ref{fig:FTIR_training}, where red lines denote Ligurian olive oil spectra. Specifically, only the former subset was previously studied \cite{Hennessy2009, Devos2014}; since the absorption of atmospheric carbon dioxide was registered in the frequency range $2400-2250$ cm$^{-1}$, whereas the end of the spectrum seemed to contain mainly noise and was thus removed too. The reason for confronting with both scenarios is twofold. On the one hand, we aim at evaluating how the presence of additional noisy variables, in an already high dimensional problem, impacts the performance of our and competing methods. On the other hand, we are interested in assessing whether a knowledge-based selection, a common practice in chemometrics studies \cite{Sato2004, Casale2006, Zou2007}, is still unavoidable even when algorithmic procedures could take over such manual approach. 

\section{Results and discussion} \label{sec:applications} 

Most classification tasks in chemometrics fall within the ``large $P$, small $N$'' family of problems: the agri-food applications considered in this paper make no exception. The direct application of model-based classifiers in such high-dimensional spaces is not possible, and typical solutions to overcome this issue include dimension reduction techniques (e.g., PCA \cite{pearson1901liii}), projection to latent structures (PLS-DA \cite{Barker2003}), single class modeling (SIMCA \cite{Wold1976}) and kernel methods (SVM \cite{Hofmann2008}). In the following, we make instead use of the stepwise REDDA procedure to provide a natural solution for dealing with corrupted high-dimensional data, and, as we will see, to identify adulterated spectra in the different scenarios.

\subsubsection*{Starches dataset} \label{sec:starches}
\begin{figure}
\includegraphics[width=\linewidth, keepaspectratio]{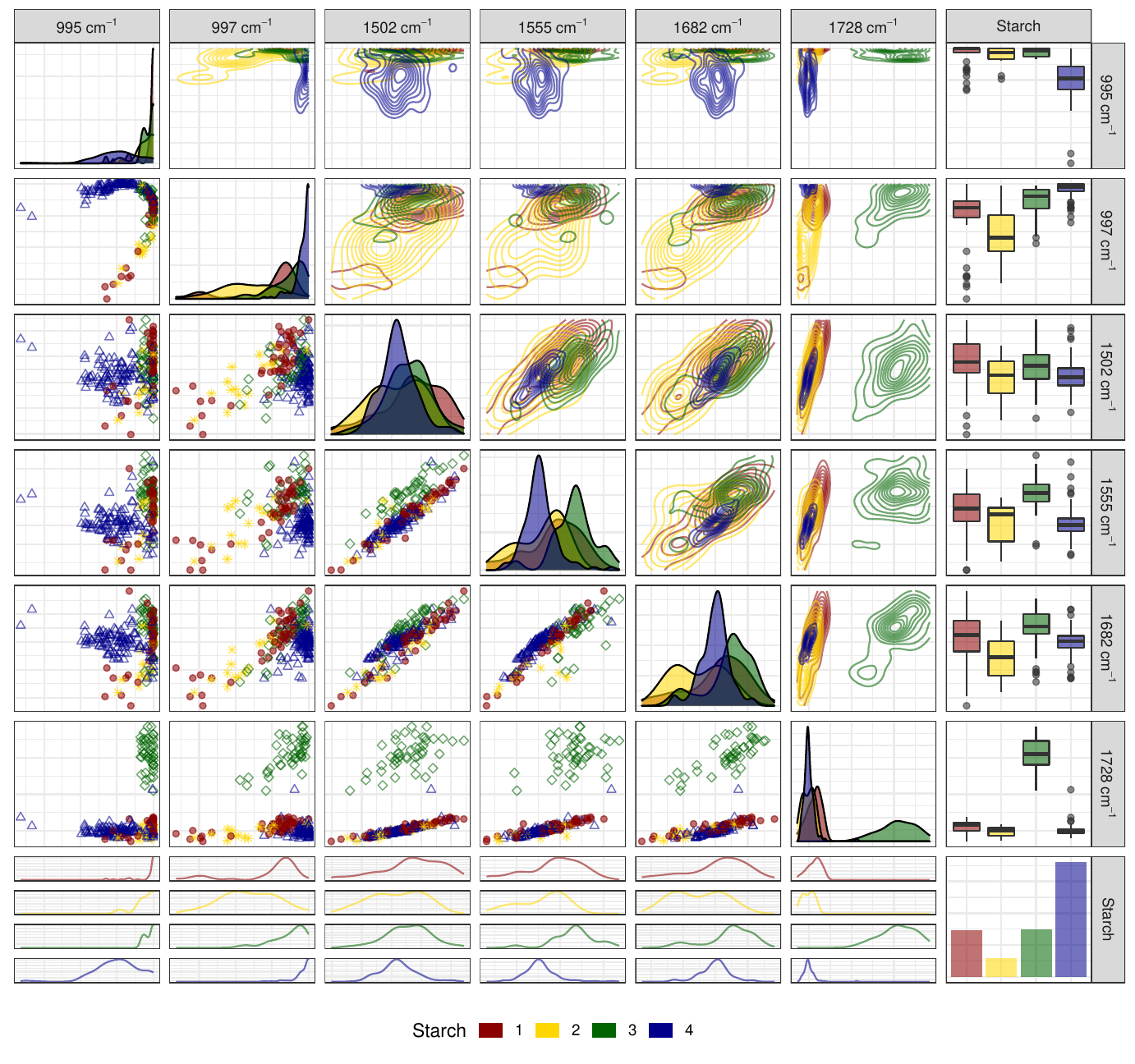}
\centering
\caption{Starches dataset: generalized pairs plot of the spectral frequencies selected by the stepwise REDDA method, training set.}
\label{fig:pair_plot_rel_var_TBIC}
 \end{figure}

For the starches dataset, we run the stepwise REDDA with $\gamma=0.03$. That is, the method is protected against potential corrupted spectra in the training set, as only $100\lceil 1-\gamma \rceil\%$ of the samples is employed for model fitting, leaving the least likely $100\gamma \%$  unmodeled. Such robustification effectively takes care of the label noise placed in the calibration set, preventing it to spoil the wavelength selection. The procedure, out of $P=2901$, selects a total of only six relevant spectral variables: $1728$ cm$^{-1}$, $1682$ cm$^{-1}$, $1555$ cm$^{-1}$, $1502$ cm$^{-1}$, $997$ cm$^{-1}$ and $995$ cm$^{-1}$. The last two wavenumbers correspond to spectral distributions of amylose and amilopectin, which are known to be present in different ratios for the different starch classes. The other wavenumbers on the list correspond to very low levels of absorbance in the spectra which makes molecular interpretation difficult. Figure \ref{fig:pair_plot_rel_var_TBIC} displays the generalized pairs plot \cite{Emerson2013} for the selected variables. Such graphical tool encompasses different plot types depending on the paired combinations of categorical and/or quantitative variables, generalizing the standard scatterplot, depicted only in the lower triangular matrix. Graphs above the main diagonal report contours of 2D density estimates, while right and bottom margins respectively include side-by-side boxplots and faceted-density plots, useful in providing a class-specific overview of univariate distributions when dealing with one categorical and one continuous variable.
Lastly, univariate plots are displayed in the main diagonal, namely density plots and a bar chart illustrating samples proportion. Figure \ref{fig:pair_plot_rel_var_TBIC} helps in visually interpreting how the selected variables contribute to classes separation. The third starch type shows on average a much higher absorbance in wavelength $1728$ cm$^{-1}$, isolating it from the other three types. The fourth starch seems to display general lower absorbance in wavelength $995$ cm$^{-1}$, while the second type exhibits a peculiar distribution in wavelength $997$ cm$^{-1}$, with much higher variability compared to the other groups. Overall, in both the bivariate scatter and in the 2D density plots it stands out that the most difficult task resides in separating starch classes $1$ and $2$, as was already pointed out in \cite{FernandezPierna2007a}.
A REDDA model with $\gamma=0.03$ is employed to classify the test samples, using as predictors the spectral frequencies retained by the stepwise variable selector. 
A Support Vector Machine with Gaussian radial kernel (SVM) was also considered, as it was shown to be the best performing classifier for this specific dataset \cite{FernandezPierna2005,FernandezPierna2007a}. In addition, we replicate the second best solution proposed by one of the `Chimiom\'{e}trie 2005' contest participants: an ensemble method was constructed by combining ROC, PLS and SVM predictions via majority vote on a subset of variables, previously determined by a PLS model. Classification accuracy for the three competing methods, learned on both the original training set and on the one containing label noise, 
are reported in Table \ref{tab:misclass_error_application}. 
\begin{table}[ht]
\centering
\caption{Number of correctly predicted test samples and associated misclassification error for different methods, starches dataset. The test set without outliers has a total sample size of $M=39$. Results with superscript $^*$ were originally reported in \cite{FernandezPierna2007a}.}
\label{tab:misclass_error_application}
\begin{tabular}{rrrr}
  \hline
 &Stepwise  &  SVM & ROC+PLS+SVM\\ 
 &REDDA&  radial kernel &\\ 
  \hline
       \multicolumn{1}{l}{Training set with label noise}   &  &  & \\ 
 $\#$ correctly predicted & 32  & 31 &31\\ 
    $\%$ correctly predicted & 82.1  & 79.5 & 79.5\\ 
    \\
           \multicolumn{1}{l}{Training set without label noise}   &  &  & \\ 
 $\#$ correctly predicted &  32 & 37$^*$ & 33\\ 
    $\%$ correctly predicted &  82.1 & 94.9$^*$ & 84.6\\ 
   \hline
\end{tabular}
\end{table}
Our robust model-based classifier attains the same predictive power when trained on either dataset: this is due to the  $\gamma$ level of trimming, that correctly identifies the adulterated spectra as to be label noise and thus safeguarding parameters from estimation bias. On the other hand, the performance of the kernel and ensemble methods are negatively impacted by the presence of the $4$ mislabeled samples. Our proposal maintains intact predictive power, further providing a great reduction in model complexity and results interpretation, being the final procedure only based on $p=6$ wavenumbers. The tremendous decrease in data dimension, together with the ability of successfully dealing and identifying label noise are two most desirable aspects in chemometrics. Overall, the selection of only six frequencies seems sufficient to well-capture the heterogeneity in the starches population. 

We mentioned at the beginning of the previous Section that $4$ adulterated spectra were manually placed in the test set (see Figure \ref{fig:starches_modifications}). While the performance of the different methods has been evaluated on the clean units only to assure fairness in the comparison, our methodology can be simultaneously employed to perform outlier detection considering the estimated marginal density for each test unit $\mathbf{y}_m$:
\begin{equation} \label{d_value_varsel}
\hat{p}(\mathbf{y}_{m,\hat{F}};\hat{\tau},\hat{\boldsymbol{\mu}}_{\hat{F}}, \hat{\boldsymbol{\Sigma}}_{\hat{F}})=\sum_{g=1}^G \hat{\tau}_g \phi \left(\mathbf{y}_{m,\hat{F}}; \hat{\boldsymbol{\mu}}_{g,\hat{F}}, \hat{\boldsymbol{\Sigma}}_{g,\hat{F}} \right)
\end{equation}
where $\hat{F}$ denotes the relevant variables identified by the stepwise REDDA approach. The $3$ spectra with lowest value of \eqref{d_value_varsel} are actually outliers. The only neglected anomaly is the one that was corrupted with a spike on a single wavelength, not identified as relevant by the feature selection method. Consequently, its marginal density in \eqref{d_value_varsel} is not altered by the manual modification. All things considered, our approach is able to effectively identify $3$ out of $4$ outliers and to greatly decrease problem complexity, whilst still maintain competitive predictive power when compared with state-of-the-art classifiers.

\subsubsection*{Meat dataset} \label{sec:meat}
\begin{figure}
\includegraphics[width=\linewidth, keepaspectratio]{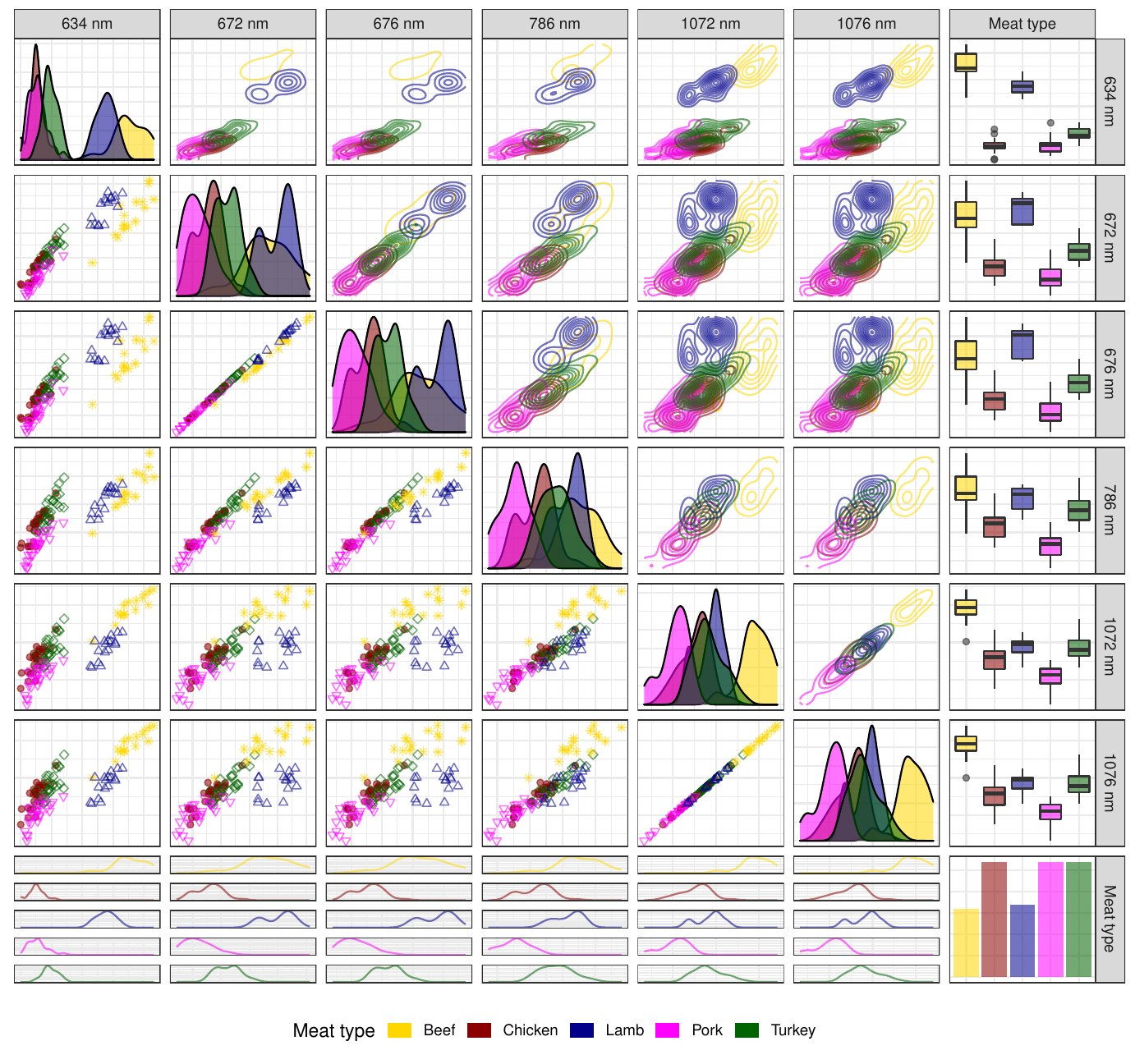}
\centering
\caption{Meat dataset: generalized pairs plot of the spectral frequencies selected by the stepwise REDDA method, test set.}
\label{fig:pair_plot_rel_var_TBIC_meat}
 \end{figure}
  \begin{table}[t]
\centering
\caption{Number of correctly predicted test samples and associated classification accuracy for different methods, meat dataset. Results with superscript $^*$, superscript $^\dagger$ and superscript $^\ddagger$ were respectively reported in \cite{Murphy2012}, \cite{Gutierrez2014} and \cite{Singh2019}.}
\label{tab:meat}
\begin{tabular}{l|cccc}
\hline
&\multicolumn{2}{c}{Training set with outliers}\:  \: &\multicolumn{2}{c}{Training set without outliers}\\
\\
&$\#$ correctly &$\%$ correctly&$\#$ correctly&$\%$ correctly\\
&predicted&predicted&predicted&predicted\\
  \hline
  \\
  Stepwise REDDA &106&  93.0&107&93.9\\
 Murphy et al \cite{Murphy2012} &- &-&107$^*$&93.9$^*$\\
 Gutierrez et al  \cite{Gutierrez2014} &-&-&-&87.4$^\dagger$\\
 PLS-DA \cite{Singh2019} &100&87.7&-&94.0$^\ddagger$\\
 VIP &102&89.5&106&93.0\\
 SR &84&73.7&105&92.11\\
   \hline
\end{tabular}
\end{table}
The stepwise REDDA method with $\gamma=0.04$ is applied to the meat dataset: the aim here is to discriminate the five meat types whilst protecting the fitting procedure from the $4$ anomalous spectra manually placed in the training set. The classification accuracy on the test samples is detailed in Table \ref{tab:meat}, where benchmark results obtained by learning the classifiers on an outlier-free training set are also reported. In addition, we compare the predictive performance of our methodology with most advanced methods (the interested reader is referred to \cite{Murphy2012, Gutierrez2014, Singh2019} for the associated classification studies) as well as standard variable selection criteria used in chemometrics, namely variable importance in projection (VIP) and selectivity ratio (SR). Results displayed in Table \ref{tab:meat} highlight that a small percentage of adulteration (the $4$ outliers account for $3.4\%$ of the training sample size) causes a reduction in terms of predictive ability for non-robust methods. Despite the decline in accuracy being modest for some procedures, we underline the advantage of dealing with the outliers within the selection algorithm: the $4$ anomalous spectra are automatically identified by the proposed approach.
Stepwise REDDA, out of $P=1050$, selects a total of six relevant wavelengths: $634$ nm, $672$ nm, $676$ nm, $786$ nm, $1072$ nm and $1076$ nm. Such wavelengths span a spectral region related to proteins, where $4$ of them belong to the visible part of the spectrum. The frequency selection is in agreement with previous results reported in the literature, for which the spectral region close to $635$ nm was found to be important when separating the red and white meat samples \cite{Liu2000,Arnalds2004}. In addition, a recent study on the identification of jowl meat adulteration in pork using hyperspectral imaging \cite{Jiang2020} acknowledges that wavenumber $676$ nm is related to the presentation of redness, corroborating our results.

The number of retained wavenumbers is much smaller compared to $407$ and $463$ obtained via VIP and SR, respectively. A generalized pairs plot is reported in Figure \ref{fig:pair_plot_rel_var_TBIC_meat}, where we observe how the spectral variables selected by our method indeed reveal distinctive meat-specific univariate and bivariate distributions, even though the poultry classes, namely chicken and turkey, are still difficult to distinguish one another. As a last worthy note, we remark that stepwise REDDA is able to cope also with outliers-free situations (see results in Table \ref{tab:meat}), nevertheless, we acknowledge that whenever noisy samples are not present in the dataset any of the competing procedures can be effectively used for identifying the most relevant wavelengths. To this extent, we are particularly interested in problems that are compatible with the assumed contaminated scenario, having argued how realistic such a circumstance is in day-to-day analyses. Hereof, the automatic tuning of the trimming level $\gamma$ remains a crucial step for all robust procedures based on impartial trimming. Several promising ideas have been recently proposed in the literature \cite{Garcia-Escudero2011, Dotto2018, Cerioli2018a, Cerioli2019, Riani2019a}, nonetheless, the wavelength selection adds a layer of difficulty to the fitting process so much so that the adaptation of existing methods to our framework is not so straightforward. Active research on the topic is currently being developed, and it will be the object of future work. A more general discussion on the problem is reported in \cite{Cappozzo2020}.



\subsubsection*{Olive Oil dataset} \label{sec:MIR}
\begin{figure}
\includegraphics[scale=.6]{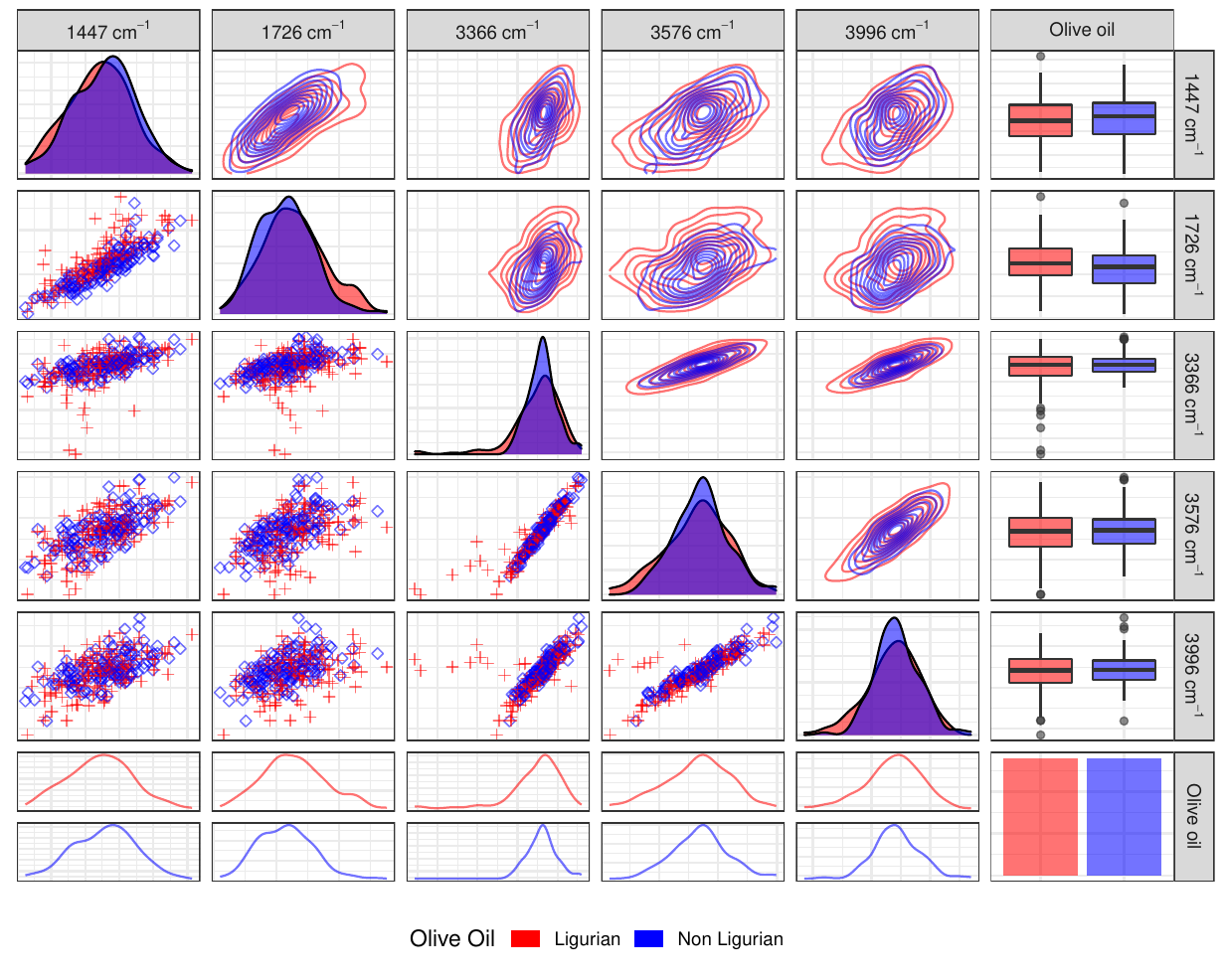}
\centering
\caption{Full olive oil dataset (frequencies $4000-700$ cm$^{-1}$): generalized pairs plot of the spectral frequencies selected by the stepwise REDDA method, training set.}
\label{fig:pair_plot_olive_full}
 \end{figure}

\begin{table}[t]
\centering
\caption{Number of correctly predicted test samples and associated classification accuracy for different methods on the two subsets of wavelengths, olive oil dataset. The test set has a total sample size of $M=630$.}
\label{tab:olive_oil}
\begin{tabular}{rrrr}
  \hline
 & Stepwise&  SVM & PLS-DA\\ 
 &REDDA&  radial kernel &\\ 
  \hline
           \multicolumn{1}{l}{Frequencies $3000-2400$ + $2250-700$ cm$^{-1}$}   &  &  & \\ 
 $\#$ correctly predicted & 507  & 459 & 509\\ 
    $\%$ correctly predicted & 80.5  & 72.9 & 80.8\\ 
        \\
           \multicolumn{1}{l}{Frequencies $4000-700$ cm$^{-1}$}   &  &  & \\ 
 $\#$ correctly predicted & 505  & 428 & 503 \\ 
    $\%$ correctly predicted & 80.2  &  67.9& 79.8\\ 
   \hline
\end{tabular}
\end{table}
\begin{figure}
\includegraphics[scale=.6]{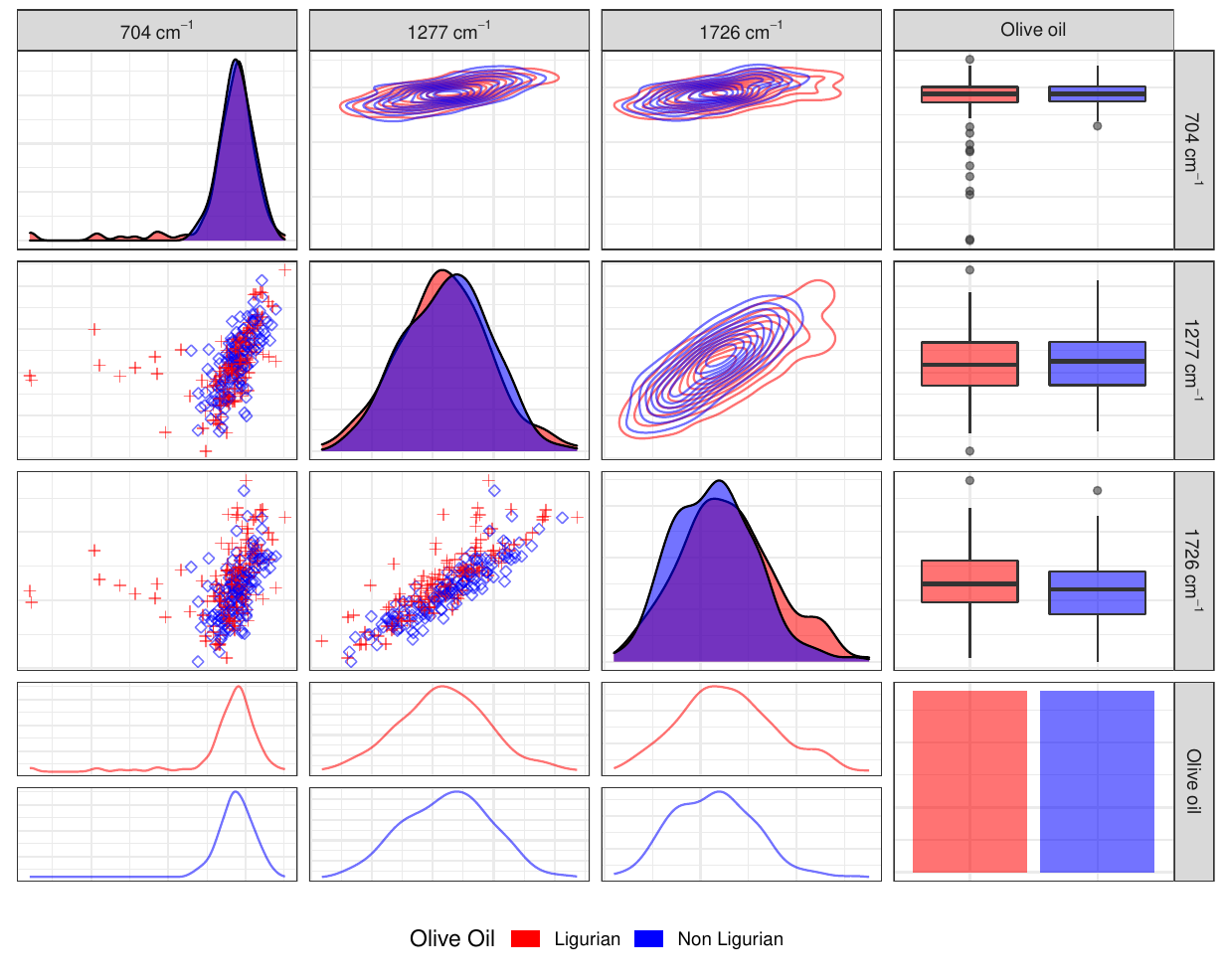}
\centering
\caption{Reduced olive oil dataset (frequencies $3000-2400$ + $2250-700$ cm$^{-1}$): generalized pairs plot of the spectral frequencies selected by the stepwise REDDA method, training set.}
\label{fig:pair_plot_olive_reduced}
 \end{figure}
\noindent Two distinct analyses, depending whether the reduced or the full spectral range (see Figure \ref{fig:FTIR_training}) is employed for model fitting, are accomplished for the olive oil dataset. Classification accuracy for both scenarios is reported in Table \ref{tab:olive_oil}, for which stepwise REDDA, partial least squares discriminant analysis (PLS-DA) and SVM classifiers have been considered. Results displayed in the table highlight some peculiarities that are worth examining. In the first place, PLS-DA and SVM are negatively impacted by the roughly $600$ more features in the full spectra case, where particularly the kernel method shows a considerable reduction in terms of predictive power. Contrarily, stepwise REDDA does not seem to be affected by the original size of the feature space, showcasing essentially unchanged classification accuracy for both spectra ranges. With reference to it, the wavenumbers selected by the  procedure amount to frequencies $704$ cm$^{-1}$, $1279$ cm$^{-1}$ and $1726$ cm$^{-1}$ when a-priori knowledge-based selection is accomplished; and to $ 1447 $ cm$^{-1}$, $ 1726 $ cm$^{-1}$, $ 3366 $ cm$^{-1}$, $ 3576 $ cm$^{-1}$ and $ 3996 $ cm$^{-1}$ in the full range scenario. In the first experiment, the relevant wavelengths correspond respectively to the $C-H$ bending of the group, to $C-C$ and $C-O$ bending situations, and to the streching of the carbonyl groups; while $ 3366 $ cm$^{-1}$, $ 3576 $ cm$^{-1}$ in the second study are associated with $O-H$ bond contributions. In both situations, there is a consistent reduction in terms of problem dimension, correspondingly moving from $1117$ and $1712$ to $3$ and $5$ retained features. Figures \ref{fig:pair_plot_olive_reduced} and \ref{fig:pair_plot_olive_full}  display the pairs plots associated to such subsets of relevant variables, in which it is apparent that this discrimination task is harder than the previous ones, with classes separation much less discernible. Particularly, no noticeable difference can be visually perceived in the univariate plots, with the only exceptions being the negative skewness of wavenumbers $704$ cm$^{-1}$ and $ 3366 $ cm$^{-1}$ for the Ligurian samples. Similarly, the bivariate plots do not strike the human eye with group-specific patterns, with only some Ligurian oil samples departing from the main point clouds. While, as expected, no wavelength was selected in the range $2400-2250$ (characterized by the uninformative absorption of atmospheric carbon dioxide), 
three ($ 3366 $ cm$^{-1}$, $ 3576 $ cm$^{-1}$ and $ 3996 $ cm$^{-1}$) out of the five variables deemed to be relevant in the full scenario were manually discarded while defining the knowledge-based reduced dataset \cite{Hennessy2009}. This unexpected result should not come as a surprise: indeed, by inspecting both the bottom panel in Figure \ref{fig:FTIR_training} and the pairs plot in Figure \ref{fig:pair_plot_olive_full}, it is evident that wavelengths greater than $3000$ cm$^{-1}$ do possess some discriminating power. Thereupon, we argue that manual based spectral range election shall be performed with care, as some valuable information may be inadvertently lost.

\section{Conclusion}
The aim of the paper has been to showcase the benefits of a robust variable selection method for classification in chemometrics.	Specifically, motivated by three agri-food applications, we have investigated the effect that contamination produces in standard tools for spectroscopic analysis, and how the proposed methodology can cope with it. Identifying noise as a factor that makes class discrimination more challenging, we have confronted label noise (starches dataset), attribute noise (meat dataset) and noisy variables (olive oil dataset). Excellent results have been obtained in all three scenarios, wherein our robust feature selection has attained a reduction in problem complexity and accurate detection of mislabeled and/or adulterated spectra, whilst maintaining competitive predictive power. In addition, we have demonstrated that our method is directly applicable to raw spectra, without needing any preprocessing step.

Mislabeling is an issue oftentimes overlooked in analytical chemistry: a method that accomplishes variable selection, eliminating the need of manual approaches, while automatically protecting against potential contamination seems particularly desirable. Furthermore, the uncovering of the most discriminative frequencies or wavenumbers both facilitates chemometrics interpretation and generates drastic cost reduction. As a consequence, laser diodes can be employed to record only targeted wavenumbers, without the need to acquire, process and store the whole spectra.

In conclusion, based on our findings, we believe that the proposed procedure could be well-accepted by the chemometric community, while additional analyses may further validate its applicability in the spectroscopic field.\\

\section*{Acknowledgments}
Andrea Cappozzo and Francesca Greselin's work is supported by Milano-Bicocca University Fund for Scientific Research,  2019-ATE-0076. Brendan Murphy's work is supported by Science Foundation Ireland grants (SFI/12/RC/2289\_P2 and 16/RC/3835). The authors are grateful to the editor and to two anonymous reviewers, whose valuable comments greatly improved the quality of the paper. 


\end{document}